\documentclass[a4paper]{jpconf}
\usepackage{graphicx}
\usepackage{amsmath}
\usepackage{amssymb}
\usepackage{amsfonts,amsthm,bm}
\usepackage{color,xcolor}
\newcommand{\bea}{\begin{eqnarray}}
\newcommand{\eea}{\end{eqnarray}}
\usepackage{cleveref}
\usepackage{here}

\begin{document}
\title{Anomalies, the Dark Universe and Matter-Antimatter asymmetry}

\author{Nick E. Mavromatos$^{1,2}$}

\address{$^1$ Department of Physics, School of Applied Mathematical and Physical Sciences, National Technical University of Athens, Zografou Campus, Athens, 15780, Greece}
\address{$^2$ Theoretical Particle Physics and Cosmology Group, Department of Physics, King's College London, Strand, London, WC2R 2LS.}

\ead{mavroman@mail.ntua.gr}

\begin{abstract}
I review a (3+1)-dimensional, string-inspired cosmological model with gravitational anomalies (of Chern-Simons (CS) type) at early epochs, and a totally-antisymmetric torsion, dual to a massless axion-like field (``gravitational axion''), which couples to the CS term. Under appropriate conditions, primordial gravitational waves can condense, leading to a condensate of the CS anomaly term. As a consequence, one obtains inflation in this theory, of running-vacuum-model (RVM) type, without the need for external inflatons. At the end of the inflationary era, chiral fermionic matter is generated, whose gravitational anomalies cancel the primordial ones. On the other hand, chiral anomalies of gauge type, which are also generated by the chiral matter, remain present during the post-inflationary epochs and become responsible for the generation of a non-perturbative mass for the torsion-related gravitational axion, which, in this way, might play the r\^ole of a Dark Matter component of geometrical origin. Moreover, in this model, stringy non-perturbative effects during the RVM inflationary phase generate periodic structures for the potential of axion-like particles that arise due to compactification, and co-exist with the gravitational axions. Such periodic potential modulations may lead to an enhanced production of primordial black holes during inflation, which in turn affects the profile of the generated gravitational waves during the radiation era, with potentially observable consequences. This model also entails an unconventional mechanism for Leptogenesis, due to Lorentz-violating backgrounds of the gravitational axions that are generated during inflation, as a consequence of the anomaly condensates, and remain undiluted in the radiation era.
\end{abstract}

\section{Introduction}\label{sec:intro}
\vspace{0.2cm}

One of the currently unsolved greatest mysteries in fundamental physics is the nature of the dark sector of the Universe, which according to fits to the plethora of the available cosmological data~\cite{Planck} corresponds to a current-epoch energy budget for the Universe, in which $\sim 70$\% is dark energy and about $\sim 26$\% Dark Matter (DM), and only a $\sim 5$\% corresponds to the known matter (mostly baryonic). The fits to the data have been made within the framework of a de-Sitter spacetime, with a a positive cosmological constant $\Lambda > 0$, in which there is a Cold-Dark-Matter (CDM) component, the so-called $\Lambda$CDM framework. 

Although at large scales, such $\Lambda$CDM-based fits prove to be quite successful, nonetheless there are important issues concerning the microscopic foundations of $\Lambda$CDM, most importantly its compatibility with UltraViolet (UV) complete models of quantum gravity (QG), which low-energy cosmologies could be embedded to~\cite{Swamp1,Swamp2,Swamp3,Palti}, but also recently its small-scale phenomenology, in particular tensions in the data~\cite{tensions}, concerning a discrepancy between the value of the current-era Hubble parameter $H_0$ inferred by direct local measurements of nearby galaxies and that based on $\Lambda$CDM fits ($H_0$ tension), but also tensions concerning the growth of structures in the Universe ($\sigma_8$ tension). Such tensions, although still susceptible to more mundane explanations, {\it e.g.} they could be due to statistical and other astrophysical uncertainties~\cite{Freedman}, nonetheless, due to their persisting nature, they prompted scientists to look for alternative to $\Lambda$CDM models as potential explanations~\cite{DiValentino}.

An interesting framework that provides a potential alleviation of both tensions ($H_0$ and $\sigma_8$~\cite{bothRVM}), but also leads to a smooth description of the evolution of the Universe from a dynamically-induced inflationary era, without the need for external inflatons, to the present era~\cite{Lima1,Lima2,Lima3},  including an explanation of the Universe thermal history~\cite{Limatherm1,Limatherm2,Solatherm},
is the so-called Running Vacuum Model (RVM)~\cite{Sola1,Sola2,Sola3,Sola4}. According to the latter, the equation of state (EoS) of the ground state of the evolving Universe  is that of a de Sitter spacetime, 
\bea\label{eosrvm}
\rho_{\rm vac} = - p_{\rm vac}\,, 
\eea
where $\rho_{\rm vac}$ ($p_{\rm vac}$) is the vacuum energy density  (pressure), but there is no cosmological constant. Instead the energy density, due to reasons of general covariance, is a function of (even) powers of the Hubble parameter $H(t)$, which evolves with the cosmic time $t$:
\bea\label{rvmenerden}
\rho_{\rm vac} (H(t))= \frac{3}{\kappa^2} \Big( c_0 + \nu \, H(t)^2 + \alpha_I \, \frac{H(t)^4}{H_I} + \dots \Big)\,,
\eea
with $\kappa = M_{\rm Pl}^{-1}$ the gravitational constant in (3+1)-dimensions, and $M_{\rm Pl}=2.435 \times 10^{18}$~GeV the reduced Planck mass scale. The quantities $c_0, \, \nu, \, \alpha_I$ are dimensionless constants, 
and $H_I \sim 10^{-5}\, M_{\rm Pl}$ is the inflationary scale, as measured by recent data~\cite{Planck}. 
The equation (\ref{rvmenerden}) was initially conjectured in \cite{Sola1,Sola2,Sola3}, based on a renormalization-group (RG) interpretation of the cosmic evolution, in which  $\rm ln (H)$ plays the r\^ole of a RG scale, and $d\rho/{\rm ln}H$ is the analogue of a RG $\beta$ function. From this point of view, the constant $c_0$ is an arbitrary constant, which would play the r\^ole of a cosmological constant in the current era. 
It worths noticing that from a phenomenological point of view, the entire history of the Universe can be described by the terms exhibited explicitly on the right-hand side of (\ref{rvmenerden}). 
At early epochs, the non-linear $H^4$ terms are dominant and lead to dynamical inflation without the need for external inflatons~\cite{Lima1,Lima2}. 

The RVM vacuum expansion (\ref{rvmenerden}), together with the EoS (\ref{eosrvm}) have been explicitly verified within the context of a string-inspired cosmological model with gravitational anomalies and torsion at early epochs, proposed in \cite{bms1,bms2,ms1,ms2}, which will be the topic of discussion in the present review.
In particular, primordial gravitational waves (GW), which could themselves have been produced in a pre-inflationary era due to various mechanisms suggested in \cite{ms1}, can lead to condensates of gravitational anomaly terms the model possesses, which in turn lead to terms $H^4$ in the cosmological vacuum energy density  of RVM type (\ref{rvmenerden}).  It is important to note that in the string context, there is no cosmological constant $c_0$ term, though, in agreement with the swampland criteria~\cite{Swamp1,Swamp2,Swamp3,Palti}, but also perturbative scattering-matrix arguments~\cite{scatt1,scatt2}, which imply that a non-zero $c_0$ constant term  would be in conflict with the UV completion (UV) of the string-inspired cosmology by the underlying microscopic string theory, or more general (a yet to be determined) quantum-gravity (QG) theory. In the model of \cite{bms1,bms2,ms1,ms2} during inflation $c_0=0$, while it has been argued~\cite{bms1} that in modern eras a
{\it metastable} de Sitter spacetime may arise, e.g. as a result of some condensate of cosmic electromagnetic fields. 

In more recent works~\cite{Solaqft1,Solaqft2,Solaqft3}, such an expansion, but with the $H^4$ terms having been replaced by $H^6$ (and higher order) terms, has been derived in the context of a novel renormalisation approach to the stress tensor of quantum scalar fields in curved space times. 
In such an approach, in agreement with the aforementioned incompatibility of string theory with de Sitter spacetimes, the constant $c_0$ is not a constant, but a slowly evolving function of cosmic time, which arises as a result of the specific techniques used in the absorption of UV infinities in the expression for the stress-energy tensor of the theory. Moreover, in this framework, 
as a consequence of the back reaction of the matter fields onto the background spacetime geometry, which is assumed to be a Robertson-Walker expanding Universe spacespacetimetime, the equation of  state of the system is different from (\ref{eosrvm})~\cite{Solaqft3}. 

Nonetheless, integrating out graviton fields in a path integral, which is essential in any consistent treatment of deriving cosmologies from string or any other UV complete QG theory, will result~\cite{Houston1,Houston2,bmssugra} in the appearance of non-polynomial terms, $H^m\,{\rm ln} (H)$, $m=2,4,\dots$, in (\ref{rvmenerden}). 
Such terms, which are due to QG corrections, may contribute to the current-era phenomenology of such models, in particular to the alleviation of the cosmological-data tensions. A detailed analysis and comparison with the results of \cite{bothRVM} within the traditional RVM framework, is in progress~\cite{gvms}. Such terms will not affect qualitatively or quantitatively the arguments of \cite{bms1,bms2,ms1,ms2} on the anomaly-condensate-induced RVM inflation. We also note that  $H^2\,{\rm ln} (H)$ structures also characterise the (renormalised) quantum field theories in curved space time of \cite{Solaqft1,Solaqft2,Solaqft3}, but there are some differences from the QG models, which may be phenomenologically distinguishable~\cite{gvms}.

The structure of the talk is the following: in the next section \ref{sec:anom}, we discuss the potentially important r\^ole of (gravitational) anomalies and torsion on inducing RVM-type inflation in the string-inspired cosmological model, without the need for external inflatons. In section \ref{sec:pbh}, we discuss the r\^ole of periodic modulations of the potential of axion fields arising from compactification in  the underlying microscopic string theory model on inducing an enhanced density of primordial black holes during inflation, which in turn will affect the profile of GW during the radiation era~\cite{Stamou}, with in principle observable consequences in future interferometers like LISA~\cite{LISA,Fumagalli}. In section \ref{sec:maasym}, we explain how, in the above anomalous class of string-inspired models, one may generate a matter-antimatter asymmetry in the Universe as a consequence of torsion-induced axion backgrounds which violate spontaneously Lorentz and CPT symmetries. Finally, section \ref{sec:concl} contains our conclusions and outlook.

\section{Anomalies and Running-Vacuum-Model-type Inflation}\label{sec:anom}
\vspace{0.2cm}

As discussed in detail in \cite{bms1,bms2,ms1,ms2}, the basic assumption in the model is that at early epochs, only fields that exist in the massless bosonic gravitational multiplet of the string appear in the effective action, as external (thus classical) fields. Chiral fermionic matter, as well as all the other (non-gravitational) fields of the theory, are generated at the end of the RVM inflation, as a result of the decay of the RVM vacuum. 

In string theory~\cite{string}, the massless gravitational multiplet consists of (spin-zero, scalar) dilatons, $\Phi(x)$, (spin-two) gravitons, $g_{\mu\nu}(x)$, and (spin-one) Kalb-Ramond (KR) antisymmetric tensor fields, $B_{\mu\nu} (x) =-B_{\nu\mu}(x) $. Due to a U(1) gauge symmetry of the closed string sector, $B_{\mu\nu} \to B_{\mu\nu} + \partial_\mu \theta_\nu - \partial_\nu \theta_\mu$, the effective low-energy target-spacetime action  of the massless gravitational mutliplet depends only on the field strength of the KR field, $H_{\mu\nu\rho} = \partial_{[\mu} B_{\nu\rho]}$, where $[\dots]$ denotes full antisymmetrisation of the respective indices. After string compactification to (3+1)-dimensions, the effective action reads:\footnote{We follow the conventions $(+,-,-,-)$ for the metric signature, and the Riemann curvature tensor definition: $R^\lambda_{\,\,\,\,\mu \nu \sigma} = \partial_\nu \, \Gamma^\lambda_{\,\,\mu\sigma} + \Gamma^\rho_{\,\, \mu\sigma} \, \Gamma^\lambda_{\,\, \rho\nu} - (\nu \leftrightarrow \sigma)$, with Ricci tensor $R_{\mu\nu} = R^\lambda_{\,\,\,\,\mu \lambda \nu}$, and Ricci scalar $R = R_{\mu\nu}g^{\mu\nu}$.}
\bea\label{sea}
S= \int d^4 x\, \Big(-\frac{1}{2\kappa^2} \, R - \frac{1}{6\, \kappa^2} e^{-4\Phi}\, H_{\mu\nu\rho}\, H^{\mu\nu\rho} + 2 \partial_\mu \Phi \, \partial^\mu \Phi + \dots \Big) \,,
\eea
where $\dots$ denote higher-derivative terms, which are subleading in a low-energy expansion (in terms of the effective (3+1)-dimensional reduced Planck energy scale, $M_{\rm Pl} = \kappa^{-1}$ (in units $\hbar=c=1$ we work throughout)). As already mentioned, according to the basic assumption of \cite{bms1,bms2,ms1,ms2}, the action (\ref{sea}) characterises the early epoch of the string Universe. The $H^2$ terms in (\ref{sea}) can be absorbed in a generalised curvature scheme, $\overline R (\overline \Gamma)$, corresponding to a geometry with (totally antisymmetric) torsion $H_{\mu\nu\rho}$, that is corresponding to a generalised spin connection: 
\bea\label{torsionconn}
\overline \Gamma_{\,\,\,\nu\rho}^\mu = \Gamma_{\,\,\,\nu\rho}^\mu + \frac{1}{\sqrt{3}} \, H^\mu_{\,\,\nu\rho} \ne \overline \Gamma_{\,\,\,\rho\nu}^\mu \,,
\eea
where $\Gamma_{\,\,\,\nu\rho}^\mu=\Gamma_{\,\,\,\rho\nu}^\mu$ is the torsion-free Christoffel connection, and 
$\frac{1}{\sqrt{3}} \, H^\mu_{\,\,\,\nu\rho} = - \frac{1}{\sqrt{3}} \, H^\mu_{\,\,\,\rho\nu} $ is the contorsion~\cite{torsion}.

In what follows, we shall consider a constant dilaton configuration, $\Phi=\Phi_0 = {\rm constant} $, 
where the constant can be set to $0$, without loss of generality. Such a solution is 
assumed to arise self consistently from the stabilisation of $\Phi$ through some appropriate string-theory-induced potential~\cite{bms2}.
Then, on taking into account the modification of $H_{\mu\nu\rho}$ by the Chern Simons (CS) gravitational terms, which are required by the Green-Schwarz mechanism~\cite{gs} for anomaly cancellation in strings, and implementing the associated Bianchi identity for the curl of $H_{\mu\nu\rho}$ via a pseudoscalar (axion-like) Lagrange multiplier field $b(x)$ (termed KR or gravitational axion)~\cite{kaloper,svrcek}, the effective action \eqref{sea} can be expressed as:
\begin{align}\label{sea4}
S^{\rm eff}_B &=\; \int d^{4}x\sqrt{-g}\Big[ -\dfrac{1}{2\kappa^{2}}\, R + \frac{1}{2}\, \partial_\mu b \, \partial^\mu b +  \sqrt{\frac{2}{3}} \, \frac{\alpha^\prime}{96\, \kappa} \, b(x) \, R_{\mu\nu\rho\sigma}\, \widetilde R^{\mu\nu\rho\sigma}  + \dots \Big] \nonumber \\
&=\; \int d^{4}x\, \sqrt{-g}\Big[ -\dfrac{1}{2\kappa^{2}}\, R + \frac{1}{2}\, \partial_\mu b \, \partial^\mu b  -
 \sqrt{\frac{2}{3}}\,
\frac{\alpha^\prime}{96 \, \kappa} \, \partial_\mu b(x) \, {\mathcal K}^\mu + \dots \Big] \,,
\end{align}
where $\widetilde{(\dots)}$ denotes the usual curved-spacetime dual tensor~\cite{bms1}, and, in the second equality, we took into account that the CS gravitational anomaly term is a total derivative of the topological current $\mathcal K^\mu$, $\mu=0, \dots 3$. Notice, as already mentioned, that there is no cosmological constant term appearing in the effective action, in agreement with the swampland criteria for embedding the low-energy theory into an UV complete string theory~\cite{Swamp1,Swamp2,Swamp3,Palti}, or for consistency with perturbative string theory, which requires a well-defined scattering matrix~\cite{scatt1,scatt2}. The action \eqref{sea4} is a typical action of a CS gravity~\cite{jackiw,yunes}, with a massless KR axion-like field  $b(x)$. The coefficient of the CS-KR-axion interaction term, that is the inverse of the KR axion coupling, $f_b$, is fixed by the underlying string theory: 
\begin{align}\label{fbdef}
f_b^{-1} \equiv  \sqrt{\frac{2}{3}} \, \frac{\alpha^\prime}{96\, \kappa} \,, 
\end{align}
where $\alpha^\prime = M_s^{-2}$ is the Regge slope of the string, with $M_s$ the string mass scale, that may be different from  $M_{\rm Pl}$. The $M_s$ may play the r\^ole of an UV cutoff of the effective low-energy gravitational theory.

For completeness we note that, classically, the duality between $b(x)$ and the torsion three-form field $H_{\mu\nu\rho}$ may be expressed via the relation~\cite{kaloper,svrcek}:
\begin{align}\label{dualHb}
-3\sqrt{2} \, \partial_\sigma b(x) = \kappa^{-1} \, \sqrt{-g} \, \epsilon_{\mu\nu\rho\sigma} \, H^{\mu\nu\rho} \, ,
\end{align}
where $\epsilon_{\mu\nu\rho\sigma}$ is the Levi-Civita symbol, such that $\epsilon_{0123} = +1$ {\it etc.} This relation corresponds to saddle points in the path integral over the torsion field $H_{\mu\nu\rho}$ of the action \eqref{sea} for $\Phi=0$, after the insertion of the Bianchi identity constraint.

In \cite{ms1,ms2} we have discussed detailed ways for the production of condensates of the CS term 
\begin{align}\label{condens}
\mathcal C_{\rm CS} = \langle R_{\mu\nu\rho\sigma}\, \widetilde R^{\mu\nu\rho\sigma} \rangle\,,
\end{align}
which could be due to a condensation of either primordial GW or (rotating) primordial Black Holes (pBH), in the presence of both of which configurations the CS term is non vanishing~\cite{alexander,lyth,kaloper,jackiw,yunes}. Thus, the gravitational anomaly is unavoidable in our model, which involves only fields from the massless gravitational string multiplet~\cite{bms1,bms2,ms1,ms2} and not gauge fields that are assumed to be generated at the end of inflation (hence, we cannot have anomaly cancellation until then). 

The GW can be produced by merging of rotating pBH, or by the non-spherically-symmetric collapse of domain walls, the latter arising, e.g., in dynamically-broken supergravity (SUGRA) models that may characterise 
pre-inflationary epochs (the dynamical breaking of SUGRA could be achieved, for instance, through condensation of gravitino fields~\cite{Houston1,Houston2}; the latter being the supersymmetric partner of gravitons still belong to the initially massles gravitational multiplet of the superstring, and hence the basic assumptions of \cite{bms1,bms2,ms1,ms2} are valid under such supersymmetric extensions. The gravitinos become very massive (even acquiring masses close to Planck mass) in such scenarios~\cite{bmssugra,ms1}, and eventually decouple from the spectrum). In such models, one may face a first hill-top inflation, near the origin of the double well gravitino-condensate potential~\cite{ellisinfl}, which provides, through instabilities of the SUGRA model~\cite{ms2}, a natural explanation of the tunnelling process to the RVM inflationary metastable ground state of the cosmological model, and the homogeneity and isotropy conditions at the beginning of the RVM inflation~\cite{bms1}.

The RVM phase is induced by the non-linearities of the vacuum energy of the model, which as a consequence of, e.g., the condensation of GW, assumes the form~\cite{alexander,lyth,MavLV}:
 \begin{align}\label{condensateN2}
\langle R_{\mu\nu\rho\sigma} \, \widetilde R^{\mu\nu\rho\sigma} \rangle 
=\frac{\mathcal N(t)}{\sqrt{-g}}  \, \frac{1.1}{\pi^2} \, 
\Big(\frac{H}{M_{\rm Pl}}\Big)^3 \, \mu^4\, \frac{\dot b(t)}{M_s^{2}} \equiv n_\star \, \frac{1.1}{\pi^2} \, 
\Big(\frac{H}{M_{\rm Pl}}\Big)^3 \, \mu^4\, \frac{\dot b(t)}{M_s^{2}}~,
\end{align} 
where the overdot denotes derivative with respect to the cosmic time, and $n_\star \equiv \mathcal N(t)/\sqrt{-g}$ is the proper density of sources of GW, which we assume approximately constant during the RVM inflation for simplicity and concreteness. The quantity $\mu$ is the UV cutoff of the effective low-energy theory, which serves as an upper bound for the momenta of the graviton modes that are integrated over in the computation of the condensate \eqref{condens} in  the presence of chiral (left-right asymmetric) GW, leading to the result \eqref{condensateN2}.

In the context of the low-energy string effective field theory, it is natural to identify $\mu = M_s$ in \eqref{condensateN2}. One may then fine tune $n_\star$ with respect to $H(t)$, 
so that the anomaly equation implies an approximately constant (average) temporal component of the topological anomaly current $<\mathcal K^0>$ ({\it cf.} \eqref{sea4}), which is the dominant component in a
(approximately) de-Sitter (inflationary) homogeneous and isotropic background spacetime:
\begin{align}\label{anomevol}
\frac{d}{dt}  <\mathcal K^0>  & + 3 \, H\, <\mathcal K^0>  \,
\simeq \, n_\star \,  \frac{1.1}{\pi^2} \, 
\Big(\frac{H}{M_{\rm Pl}}\Big)^3 \, \mu^4\, \frac{\dot b(t)}{M_s^2} \, ,  \quad \dot b (t) =\sqrt{\frac{2}{3}} \frac{\alpha^\prime}{96\, \kappa}\, <\mathcal K^0>\,, \quad \alpha^\prime = \mu^{-2}\,, \nonumber \\
 <\mathcal K^0>  & \simeq {\rm constant} \quad \Rightarrow \quad n_\star^{1/4} \, \sim \,  7.6 \times \Big(\frac{M_{\rm Pl}}{H}\Big)^{1/2}\,,
\end{align}
where the relation of $\dot b$ with $<\mathcal K^0>$ is a consequence of the KR-axion equation of motion~\cite{bms1,bms2,MavLV}. If one uses the Planck cosmological data~\cite{Planck} to fix $H \simeq H_I $ during inflation to the range of values 
$H_I / M_{\rm Pl} < 10^{-5}$, then we obtain~\cite{MavLV} 
\begin{align}\label{nstarval}
n_\star \gtrsim 3.3\times 10^{13}\,, 
\end{align} 
which defines the range of values of the macroscopic number of sources (per proper volume) needed to produce an approximately constant gravitational anomaly condensate in the context of the CS cosmology of \cite{bms1,bms2,ms1,ms2}. 

The total vacuum energy density during inflation assumes an RVM form \eqref{rvmenerden}~\cite{Sola1,Sola2,Sola3,Lima1,Lima2}, as the explicit computations of \cite{bms1,ms1,MavLV} have demonstrated:
\begin{align}\label{totalenerden}
\rho^{\rm total}_{\rm vac} =  -\frac{1}{2}\, \epsilon \, M_{\rm Pl}^2\, H^2 + 4.3 \times 10^{10} \, \sqrt{\epsilon}\, \frac{|\overline b(0)|}{M_{\rm Pl}} \, H^4\,,
\end{align}
where we have parametrised the constant $\dot b$ axion background during the RVM inflation as 
\begin{align}\label{axionbackgr}
b(t)=\overline{b}(t_0) + \sqrt{2\epsilon} \, H \, (t-t_0) \, M_{\rm Pl}\,, 
\end{align}
with $t_0$ denoting the beginning of (the RVM) inflation. The assumption on the approximate de Sitter (positive-cosmological-constant) nature of the condensate, so as to lead to an approximately constant $H_I$ during inflation, requires~\cite{bms1,bms2}:
\begin{align}\label{bcb}
|\overline{b}(t_0)| \gtrsim N_e \, \sqrt{2\epsilon} \, M_{\rm Pl}\, = \mathcal O(10^2)\,\sqrt{\epsilon}\, M_{\rm Pl}\,,  
\end{align}
with $N_e= \mathcal O(60-70)$ the number of e-foldings of the RVM inflation. The reader should have noticed that, as a result of the CS gravitational anomaly terms,\footnote{We note that the variation of the gravitational CS anomaly terms in \eqref{sea4} with respect to the graviton field is non trivial, giving rise to the so-called Cotton tensor~\cite{jackiw,yunes}, which 
affects the Einstein equation: $$R^{\mu\nu} - \frac{1}{2}\, g^{\mu\nu}\, R - \sqrt{\frac{2}{3}}\, \frac{\alpha^
\prime \, \kappa}{12} \, C^{\mu\nu} = \kappa^2 \, T^{\mu\nu}\,.$$
The Cotton tensor is not covariantly conserved, $C^{\mu\nu}_{\,\,\,\,\,\,\,;\mu} = -\frac{1}{8}\,( \partial^\nu b )
R_{\mu\nu\rho\sigma}\, \widetilde R^{\mu\nu\rho\sigma}$ (where $;$ denotes the torsion-free gravitational convariant derivative), and this implies an exchange of energy between the axion and the gravitational anomaly terms., which leads to the aforementioned negative contributions of the CS anomaly to the stress-energy tensor $T^{\mu\nu}$ of the system.} the coefficient of the $H^2$ term in \eqref{totalenerden} is negative, in contrast to the standard form, current-era, RVM energy density \eqref{rvmenerden}, in which $\nu > 0$. Such negative contributions to the energy density are characteristic of higher-curvature gravities, such as Gauss Bonnet-dilaton gravity~\cite{kanti}. 
 
The EoS of the string-inspired model during this inflationary phase has been explicitly computed in
\cite{ms1} and found to coincide with the RVM EoS \eqref{eosrvm}, which justifies {\it a posteriori} calling this model {\it stringy RVM}. 

An important feature of the model is the Lorentz-Violating (LV) nature of the KR axion background \eqref{axionbackgr}, characterised by a constant $\dot b$, as a consequence of the (approximately) constant anomaly condensate \eqref{anomevol}
during the RVM inflation. The LV nature stems from the duality relation \eqref{dualHb}, which implies a constant
$\epsilon_{ijk} \, H^{ijk} $, $i,j,k=1,2,3$, that clearly breaks spontaneously the Lorentz symmetry.
Such a background survives the end of the inflationary phase, and can lead, during the radiation phase, to a LV and CPT-Violating (CPTV) Leptogenesis, according to the mechanism suggested in \cite{Sarkar1,Sarkar2,Sarkar3,Sarkar4}, as we shall discuss in section \ref{sec:maasym}. 
In the next section \ref{sec:pbh}, instead, we shall study the consequences of the presence of periodic modulations of the potential of axion fields that arise from compactification in string theory, and co-exist with the KR axions~\cite{svrcek,axiverse}.  We shall argue that such structures enhance the densities of primordial black hole (pBH) densities  during inflation, and, thus, affect the profiles of GW during the radiation era.

\section{Enhanced Primordial Black Holes and modified Gravitational Wave profiles}\label{sec:pbh}
\vspace{0.2cm}

In \cite{Stamou} we have considered one such toy string-inspired extension of the stringy RVM model, in which 
we considered the effects of one more axion field, arising from compactification, $a(x)$, with axion coupling $f_a$, in addition to the KR axion $b(x)$ that drove the RVM inflation. Below we shall review briefly the situation and the consequences for the phenomenology of pBH and GW, which could in principle be observable in future interferometers~\cite{LISA,Fumagalli}.

\begin{figure}[H]
\center
\includegraphics[width=20pc]{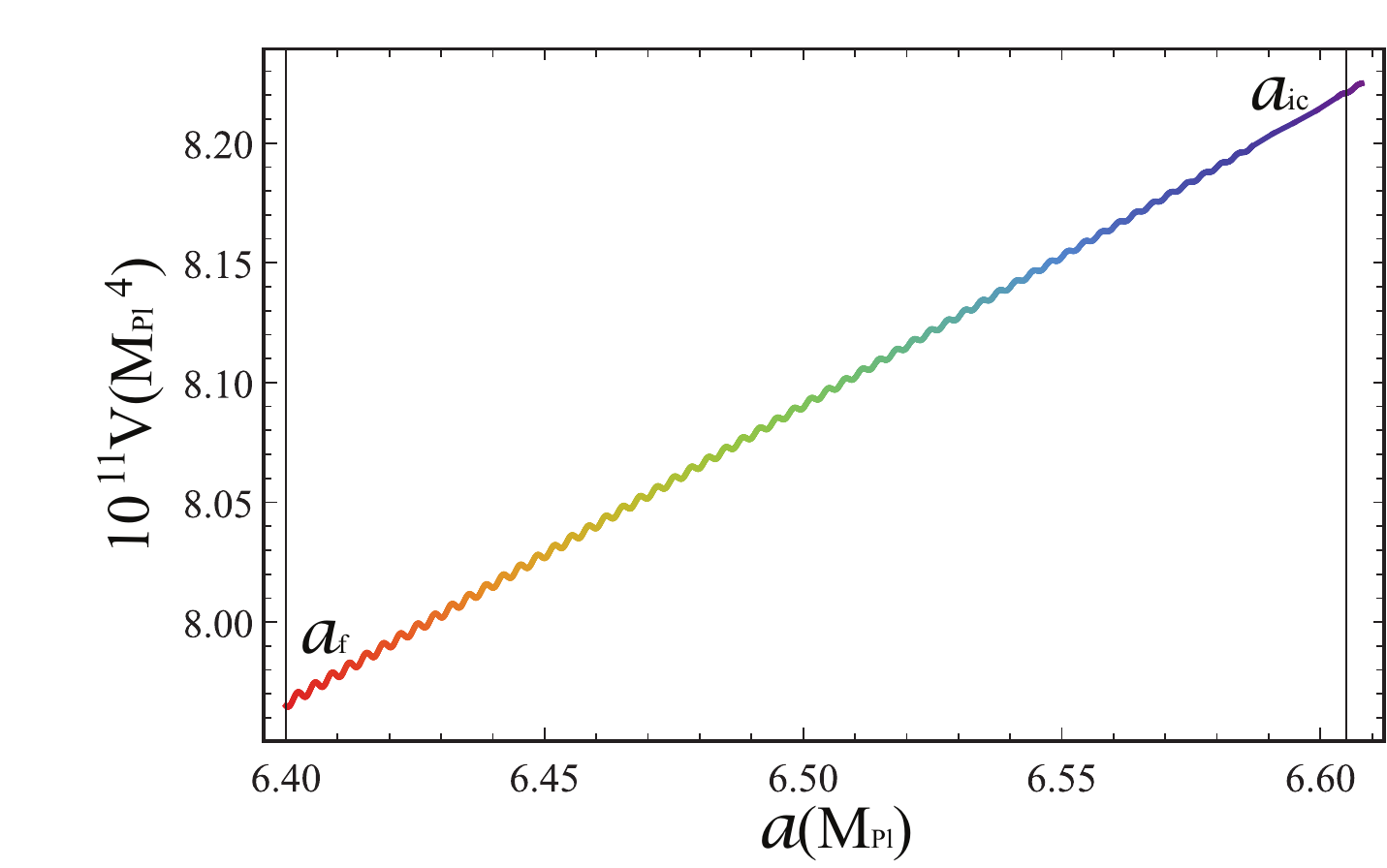}
\caption{\label{label} The  potential \eqref{effpot} in the $a$-field direction for the set of parameters
of Table \ref{table1}. $a_{ic}$ represents the initial condition for the $a$ axion, before its oscillations start to become appreciable,  and $a_{f}$ corresponds to the value of the field when the inflation ends. Picture taken from \cite{Stamou}.}
\label{fig:pot}
\end{figure}

The presence of the anomaly condensate \eqref{condens} (or \eqref{condensateN2}) leads to an approximately linear potential of the KR axion $b(x)$, but also for the axion $a(x)$, assuming a standard CS coupling of the $a(x)$ field with the gravitational anomaly~\cite{svrcek}
\begin{align}\label{axionCS}
S_{\rm a CS} = \int d^4 x \sqrt{-g}\, \frac{1}{f_a} \, a(x) \, R_{\mu\nu\rho\sigma}\, \widetilde R^{\mu\nu\rho\sigma}\,.
\end{align}
To maintain the spirit of the stringy RVM model \cite{bms1,bms2,ms1,ms2}, 
as far as the induced RVM inflation is concerned, we assume that world-sheet stringy non-perturbative effects generate dominant periodic structures for the potential of the axion field $a(x)$ only. To keep the analysis general as far as compactification effects are concerned, we also include brane compactification effects which, in the context of axion monodromy models~\cite{silver}, also lead to linear terms in the $a$-field potential, in addition to the ones due to the anomaly condensate. We assume that such effects correspond to an energy scale $\Lambda_2$. Par contrast, the gravitational anomaly effects, discussed in the previous section, are related to an energy scale $\Lambda_0$, which is defined 
through:
$ \langle R_{\mu\nu\rho\sigma}\, \widetilde R^{\mu\nu\rho\sigma} \rangle \equiv \widetilde \Lambda_0^4 \simeq \frac{1}{\pi^2}\, \sqrt{2\epsilon} \,
 \frac{\mu^4}{M_s^2\, M_{\rm Pl}^2}\, H^4 $,
where in the effective theory it is natural that we set $\mu=M_s$, as mentioned above.
Then, the corresponding condensate-induced linear axion-$b$ effective potential ({\it cf.} \eqref{sea4}) reads~\cite{Stamou} 
 \begin{align}\label{linV}
&V(b) \simeq b \, \widetilde \Lambda_0^4 \sqrt{\frac{2}{3}} \, \frac{M_{\rm Pl}}{96 \, M_s^2} = b \, \frac{\widetilde \Lambda_0^4}{f_b} \, \equiv b \, \Lambda_0^3 \,,
\end{align}
with the KR axion coupling $f_b$ given in \eqref{fbdef}. Therefore, the (3+1)-dimensional spacetime effective potential for the two axion fields, $a(x)$ and $b(x)$ that we consider for the analysis in this section acquires the form:
\begin{equation}\label{effpot}
 V(a, \, b)={\Lambda_1}^4\left( 1+ f_a^{-1}\, \tilde \xi_1 \, a(x) \right)\, \cos({f_a}^{-1} a(x))+\frac{1}{f_{a}}\Big(f_b \, {\Lambda_0}^3 + \Lambda_2^4 \Big) \, a(x) + {\Lambda_0}^3\, b(x), 
 \end{equation}
 where, in the study of \cite{Stamou}, the parameters $\tilde \xi_1, f_a$, and the world-sheet instanton induced scales $\Lambda_1, \Lambda_2 $ are treated as phenomenological, with the constraint though that~\cite{Stamou}:\footnote{For practical purposes, such a hierarchy of scales \eqref{constr} implies that the term $\Lambda_2^4$ can be ignored in front of $\Lambda^3_0 f_b$ in \eqref{effpot}.}
\begin{align}\label{constr}
 \Big(\frac{f_b}{f_a} + \frac{\Lambda^4_2}{f_a\, \Lambda^3_0} \Big)^{1/3}\, \Lambda_0  \, < \, \Lambda_1 \ll \Lambda_0 ~.
 \end{align}
 so that the RVM inflation is driven by the $b(x)$ field, and is only prolonged by the presence of the $a(x)$ field. 
 The reader should notice that the non-perturbative-world-sheet-effects-generated periodic modulations of the axion-$a(x)$ potential in \eqref{effpot} do not lead to mass terms for the axion. 
The potential \eqref{effpot} is plotted in fig.~\ref{fig:pot} for a concrete set of parameters, given in Table \ref{table1}.  
 
\begin{center}
\begin{table}[ht]
\begin{tabular}{||c| c c c c|c c c||} 
 \hline
SET & $g_1$ & $g_2$ & $\xi$ &$f(M_{Pl})$ &$\Lambda_0(M_{Pl})$ & $\Lambda_1(M_{Pl})$ &${\Lambda_3}(M_{Pl})$ \\ [0.5ex] 
 \hline\hline
 1 &$ 0.021$ & $0.904 $&$ -0.15$ & $2.5\times 10^{-4}$&$ 8.4 \times 10^{-4} $&$8.19\times 10^{-4}$ &$2.32\times 10^{-4}$\\
  \hline
\end{tabular}
\caption{An example of parameters for the potential \eqref{effpot}~\cite{Stamou}, 
where, for brevity, we use the notation: $g_1 \equiv \frac{f_b}{f_a} + \frac{\Lambda^4_2}{f_a\, \Lambda^3_0}  \,, \,
\Lambda_1^4 \equiv g_2\, \Lambda_0^4\, , \,f \equiv f_a\,, \,  \xi \equiv  \frac{M_{\rm Pl}}{f_a}\, \tilde \xi_1\,.$ }
\label{table1}
\end{table}
\end{center}

In this scenario, the KR field $b$ dominates during the first stage of (RVM) inflation, 
when the oscillations of the axion-like field $a$ are suppressed compared to its linear term.  As the cosmic time passes, the value of $b$ decreases~\cite{Stamou}, and eventually the oscillations of the  field $a$ become  the leading effect. The oscillations of the $a$ axion appear effectively as 
many tiny ``step''-like patterns, which lead to an enhancement of the primordial spectra, a feature reflected in the background of GWs, in similar spirit to the situation governing inflationary potentials with steep-like features, including discontinuities~\cite{tetradis1,tetradis2,Fumagalli2,Fumagalli3}. However, contrary to such scenarios, the evolution here is smooth, and there are no discontinuities neither in the field nor in its first derivative.

\begin{figure}[H]
\centering
\includegraphics[width=75mm,height=55mm]{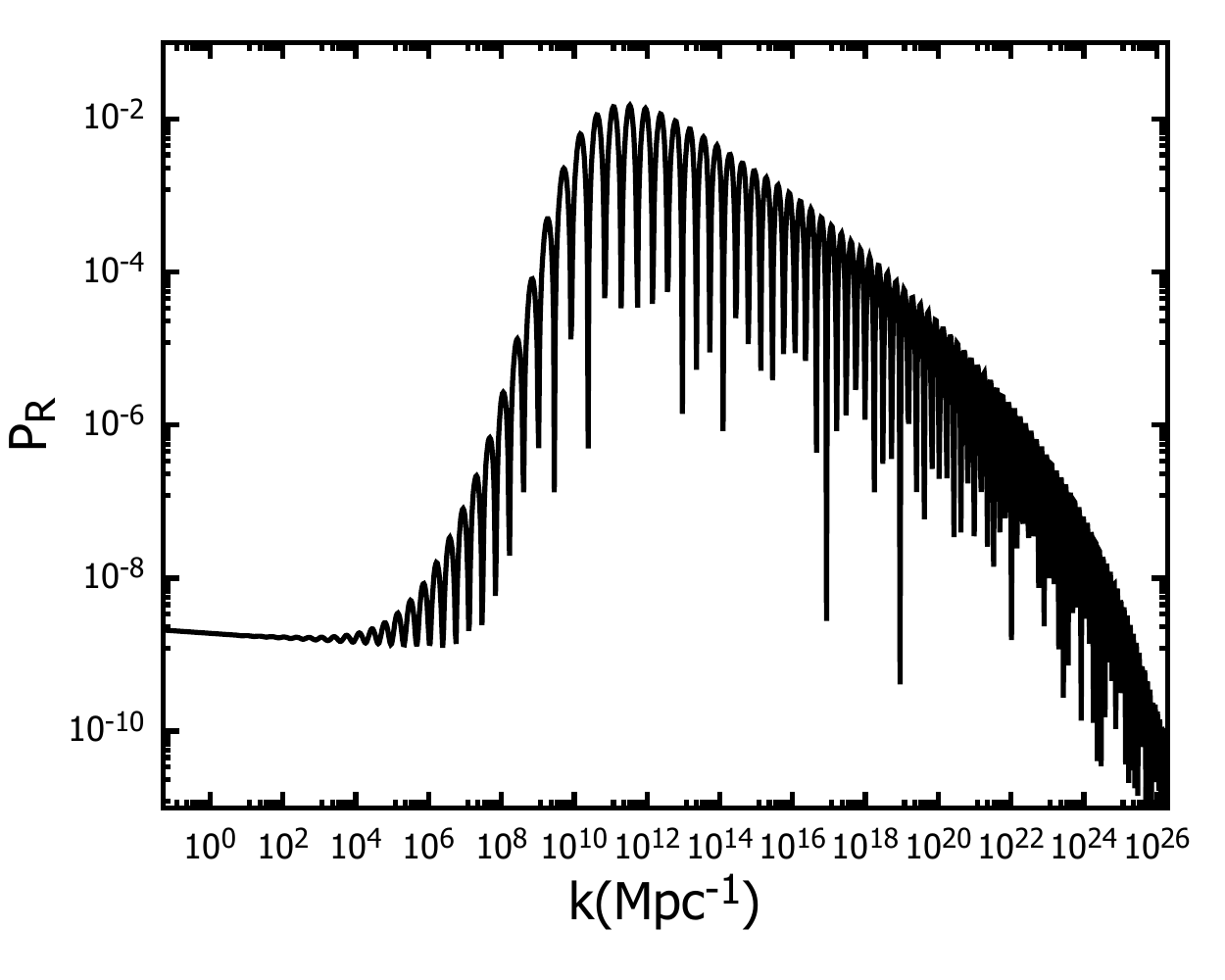}
\includegraphics[width=75mm,height= 63mm]{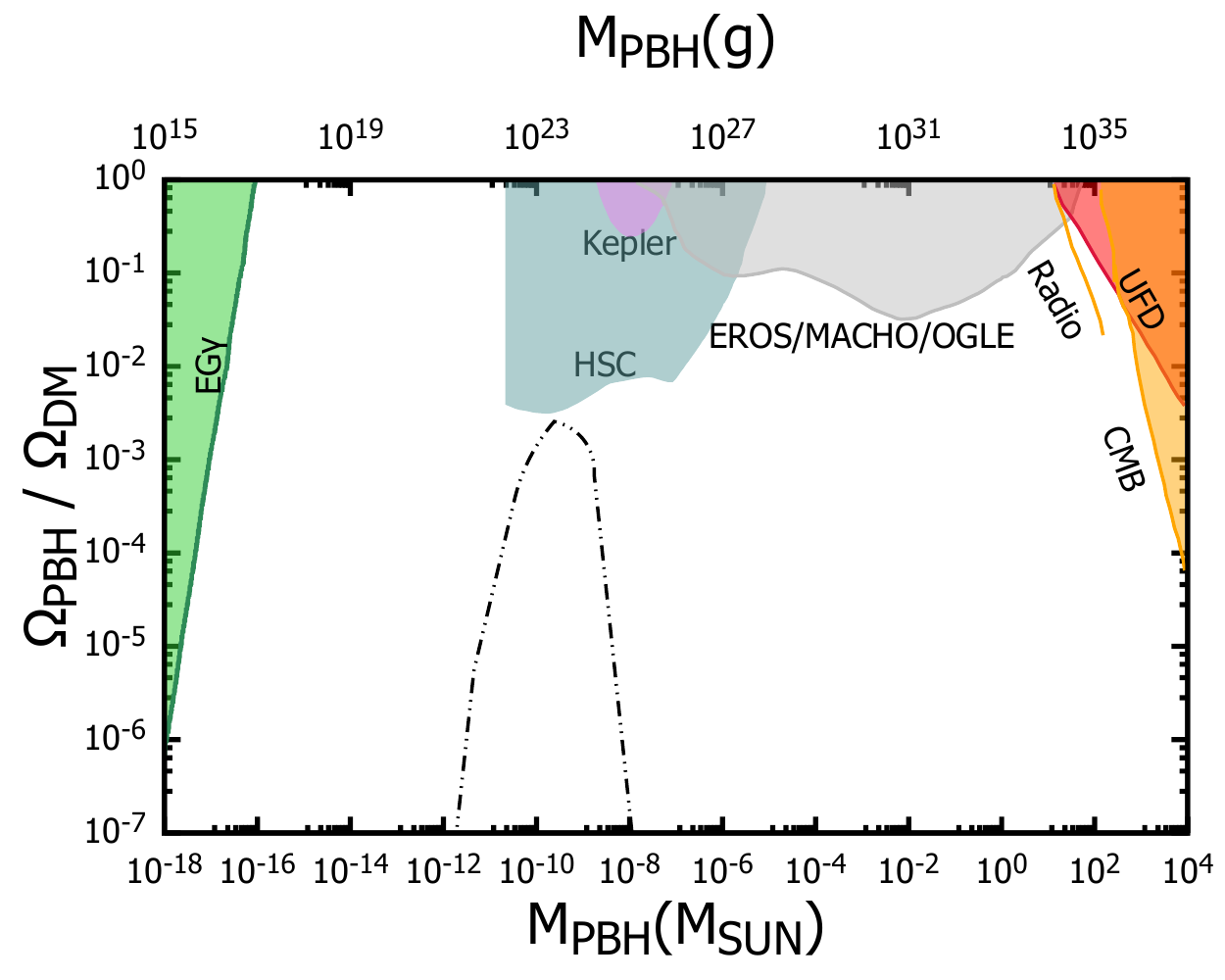} \\ \includegraphics[width=80mm]{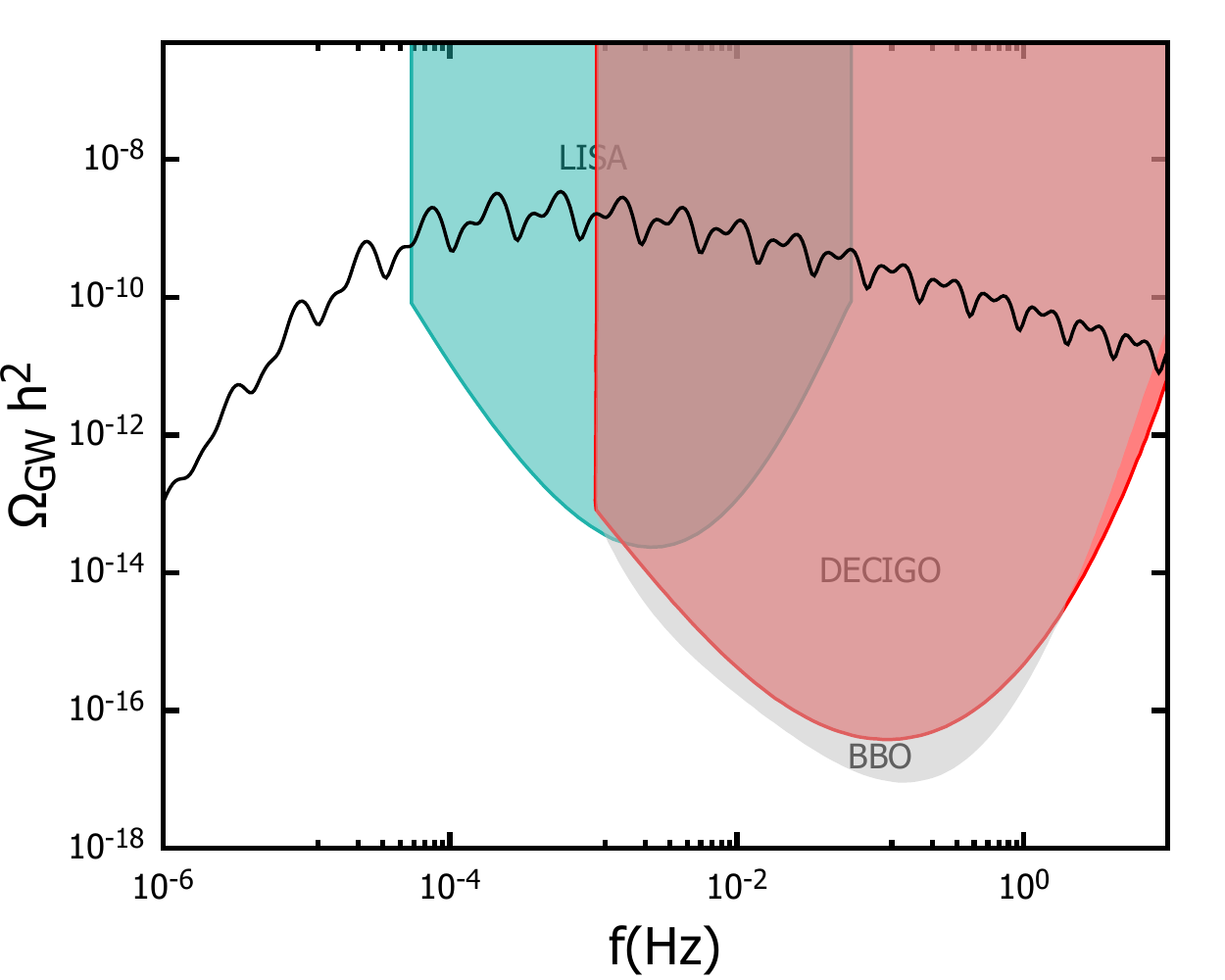}
\caption{ The results of the power spectrum, fractional abundance of PBHs and the energy density of induced GWs for the first set of parameters given in Table \ref{table1}.  The initial conditions for the axion fields are (in units of $M_{\rm Pl}$) $a_{ic}=6.605$, and $b_{ic}= 11.100 $. The fractional abundance of PBHs is  $f_ {PBH}=0.01$. Picture taken from \cite{Stamou}.}
\label{fig:power}
\end{figure}

The detailed analysis of \cite{Stamou} has demonstrated the enhancement of the power spectra due to the oscillations of the axion $a$, which in turn enhances the densities of the pBH during the RVM inflation, and consequently the profile of the GW during the radiation era that succeeds the RVM inflation~\cite{Lima1,Lima2,bms1,ms1}. This can lead to the production of a significant fraction of the DM in the form of pBHs~\cite{Carr1,Carr2,pBHDM1,pBHDM2}.
The situation is depicted in fig.~\ref{fig:power}, where we plot features of the power spectra $P_R$ in the two-axion model of \cite{Stamou}, together with the pBH abundance (as a function of pBH mass (in units of the solar mass)), as well as the energy density profile of the GW, for the set of parameters given in Table \ref{table1}. 
Similar features persist for a large range of pBH abundances $f_{PBH}$, including those of order $f_{PBH}=0.80$~\cite{Stamou}.

\begin{figure}[H]
\centering
\includegraphics[width=75mm,height= 55mm]{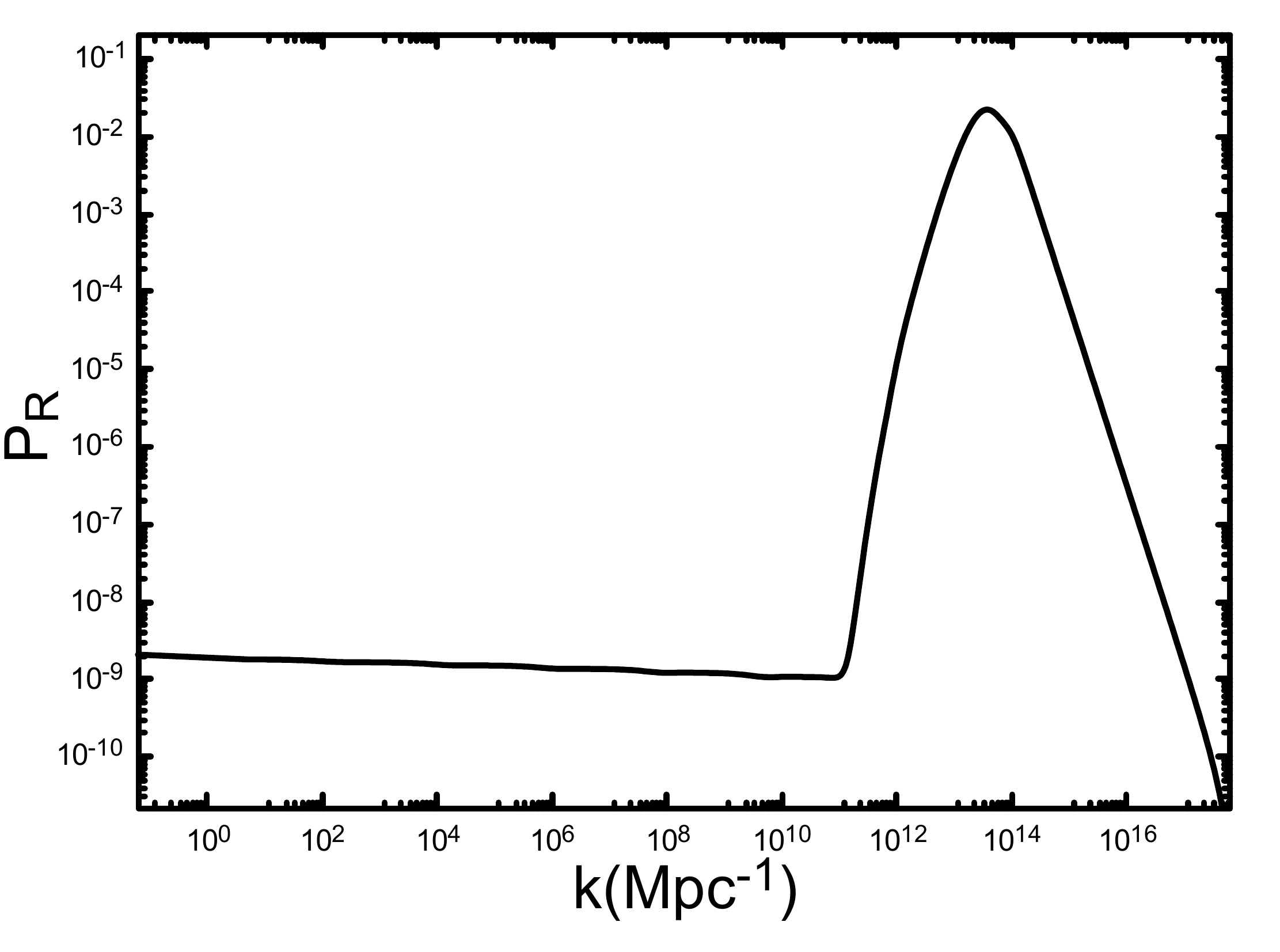}
\includegraphics[width=75mm,height= 63mm]{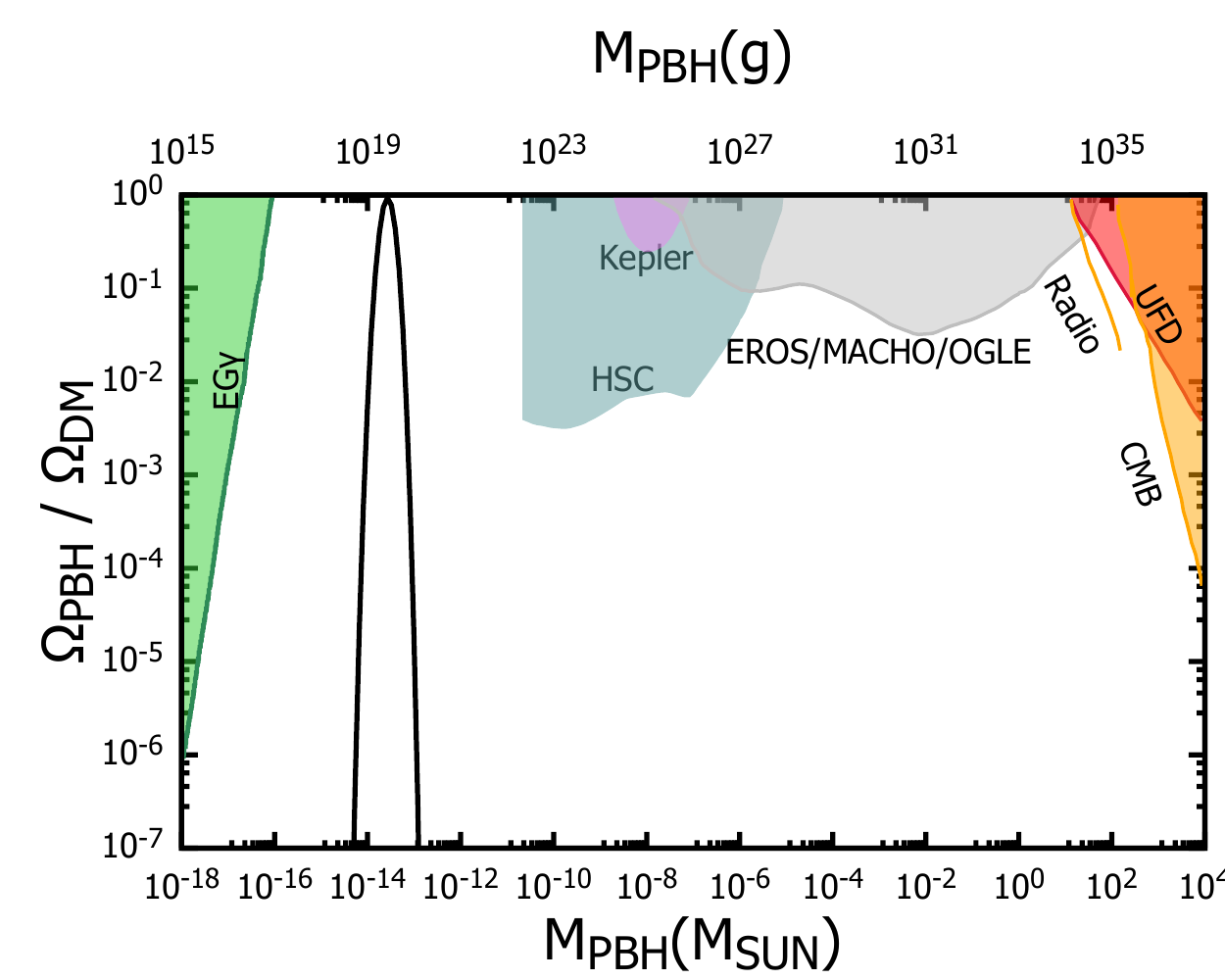}\\
\includegraphics[width=80mm]{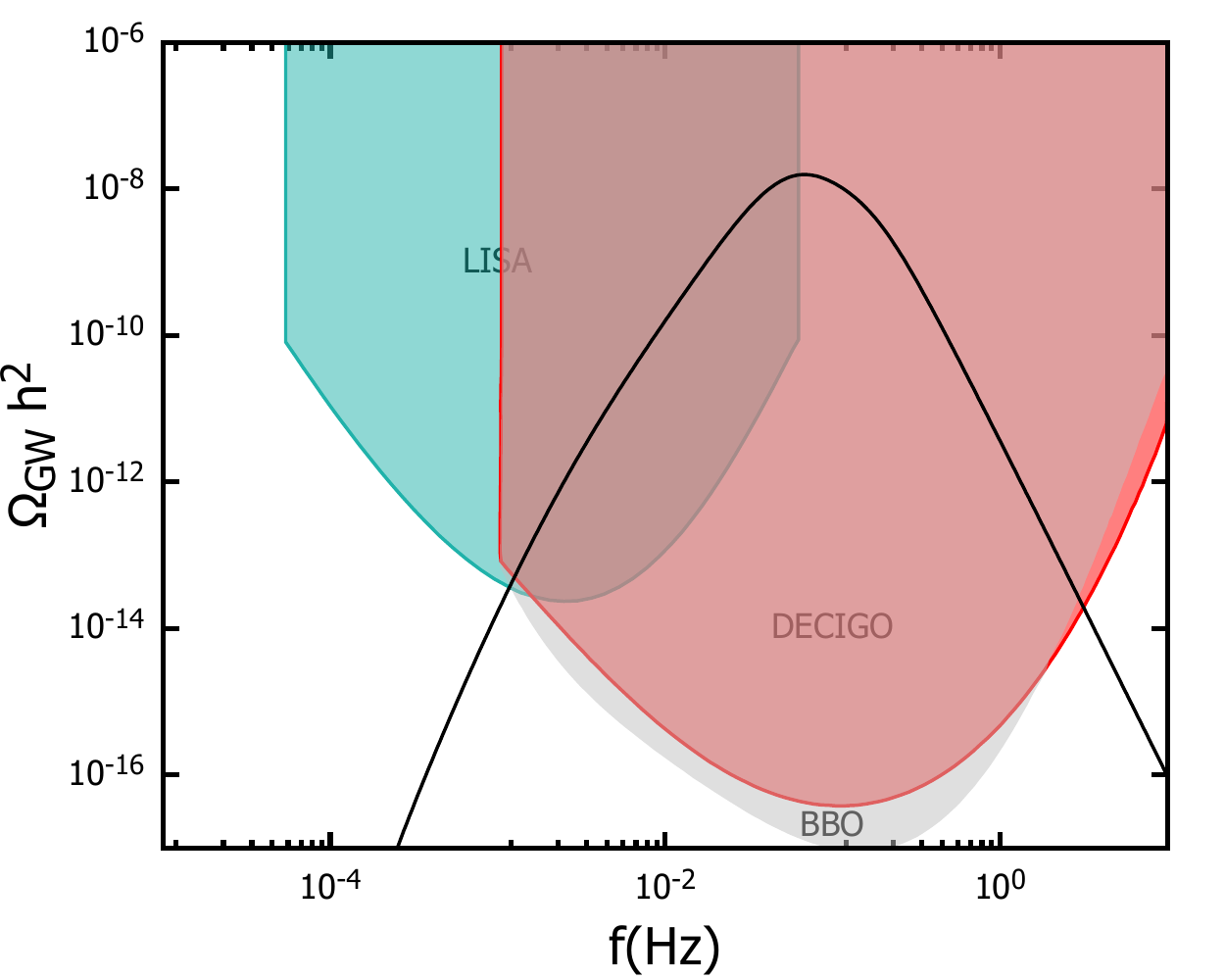}
\caption{\underline{Upper left panel}: The  power spectrum for the model of \cite{sasaki}, with the hierarchy of scales \eqref{constr3}.  \underline{Upper right panel}: The abundances of PBHs. \underline{Lower panel}: The energy density of GWs. The diagrams are given for the choice of parameters $\Lambda_0=8.4\times 10^{-4} M_{\rm Pl},  \quad  \Lambda_1=9.7 \times 10^{-3} M_{\rm Pl}\, , \,g_1=110\, , \, g_2=1.779\times
10^4\, ,\, \xi=-0.09\, , \, f=0.09 \,\, M_{\rm Pl}\,$, which obeys the hierarchy of scales \eqref{constr3}. 
The initial conditions for the axion $a$ and $b$ fields (in units of $M_{\rm Pl}$) are $(a_{ic}\, , \, b_{ic})= (7.5622, \, 0.5220)$. Picture taken from \cite{Stamou}.}
\label{fig:featureless}
\end{figure}

These results should be compared with the ones characterising other string-inspired axion-monodromy models with two axions, considered in \cite{sasaki}, which, although leading also to enhanced power spectra and enhanced production of pBH during inflation, nonetheless are characterised by a different hierarchy of scales than \eqref{constr}:
 \begin{align}\label{constr3}
 \Lambda_0 \, \ll  \, \Big(\frac{f_b}{f_a} + \frac{\Lambda^4_2}{f_a\, \Lambda^3_0} \Big)^{1/3}\, \Lambda_0  \, < \, \Lambda_1 ~, 
 \end{align}
and are such that it is the axion $a$ that drives initially inflation, whilst the KR axion $b$ is responsible for prolonging it. As becomes clear from fig.~\ref{fig:featureless}, the smooth (featureless) energy density profile of GW after inflation in that case, in contrast to the model of \cite{Stamou} examined above ({\it cf.} fig.~\ref{fig:power}), makes the models of \cite{sasaki} in principle distinguishable experimentally from the model of \cite{Stamou}, e.g. in future interferometers~\cite{LISA,Fumagalli}. 

Before closing this section, we would like to remark~\cite{Stamou} that in the context of RVM inflationary models with several axions, we may expect further enhancement of the power spectra and the densities of pBH, due to a prolonged reheating period that characterises (depending on the parameters) some of the RVM models at the exit from RVM inflation~\cite{Lima1,Lima2}.
A complete analysis of such models and a study of the associated consequences for the profile of GW after inflation is pending. 

\section{Lorentz-violation-induced Matter-Antimatter Asymmetry}\label{sec:maasym}
\vspace{0.2cm}

In this section we examine the consequences of the LV axion background \eqref{axionbackgr} for inducing unconventional LV and CPTV processes for lepton asymmetry generation (Leptogenesis) during the radiation era, in models involving massive right-handed neutrinos (RHN), within the context  
of the mechanism suggested in \cite{Sarkar1,Sarkar2,Sarkar3,Sarkar4}.

First let us review this mechanism, and then see how it can be adapted to our stringy RVM cosmological model.
We consider the following fermion action in an LV and CPTV violating axial background $b_\mu$
in the presence of massive right-handed neutrinos (RHN), with standard portals coupling the RHN sector to Standard Model (SM) lepton and Higgs sectors:
\begin{align}\label{smelag}
\mathcal{L}= {\mathcal L}_{\rm SM} + i\overline{N}\, \gamma^\mu\, \partial_\mu \, N-\frac{m_N}{2}(\overline{N^{c}}N+\overline{N}N^{c})-\overline{N}\gamma^\mu\, b_\mu \, \gamma^{5}N-\sum_f \, y_{f}\overline{L}_{f}\tilde{\phi}^dN+ {\rm h.c.}
\end{align}
where h.c.  denotes hermitian conjugate, ${\mathcal L}_{\rm SM}$ is the Standard Model (SM)  Lagrangian, $N$ is the RHN field (with $N^c$  its charge conjugate field), of (Majorana) mass $m_N$,  $\tilde \phi$ denotes the SU(2) adjoint of the Higgs field  $\phi$ ($\tilde{\phi}^d_i \equiv \varepsilon_{ij}\phi_j~, \, i,j=1,2,$ SU(2) indices), and $L_{f}$ is a lepton doublet of the SM sector, with $f$ a generation index, $f=e, \mu, \tau$, assuming three SM generations; $y_f$ is a Yukawa coupling, which, if non-zero, provides a non-trivial (``Higgs portal'') interaction between the RHN and the SM sector, used in the seesaw mechanism for the generation of SM neutrino masses~\cite{seesaw1,Yanagida,GellMann,seesaw2,seesaw2,seesaw3,seesaw4}. The LV and CPTV axial background $b_\mu$ is of Standard Model Extension (SME) type~\cite{sme1,smebounds}, and it has only a constant, temporal component~\cite{Sarkar1}
\begin{align}\label{background}
b_\mu = M_{\rm Pl}^{-1} \, \dot{b} \, \delta_{\mu\,0}\,, \quad  \mu=0, \dots 3\,, \quad \dot{\overline b} = {\rm constant}\,.
\end{align}
One then computes in that model, the tree-level decays of (even one) flavour of a massive RHN, $N$, of mass $m_N$, into SM particles and antiparticles, that is one computes the RHN decay rates in the channels: ${\rm Channel} ~I: \,  N \rightarrow l^{-}h^{+}~, ~ \nu \, h^{0}$,  and 
${\rm Channel ~II}: \, N \rightarrow l^{+}h^{-}~,~  \overline \nu \, h^{0}$,
where $\ell^\pm$ are charged leptons, $\nu$ ($\overline \nu$) are the light, "active", neutrinos (antineutrinos) in the SM sector, $h^0$ is the neutral Higgs field, and
 $h^\pm$ are the charged Higgs fields (which, at high temperatures, above the spontaneous electroweak symmetry breaking, of interest in this scenario~\cite{Sarkar1,Sarkar2,Sarkar3,Sarkar4,ms1}, do not decouple from the physical spectrum).  As a result of the non-trivial $b_0 \ne 0$ background (\ref{background}), the decay rates of the Majorana RHN between the channels I and II are different, resulting in a Lepton-number asymmetry.  
In the limit of a weak background \eqref{background} (compared to the mass of the RHN, {\it i.e.} $|b_0 | \ll m_N$) in the model \eqref{smelag}, this lepton asymmetry is found proportional to the ratio $b_0/m_N$. Assuming a freezeout temperature $T_D$ for the RHN $T_D \simeq m_N$, one finds (using semi-analytic techniques to compute the Lepton asymmetry $\Delta L$)~\cite{Sarkar2}:
\begin{align}\label{dltd}
\frac{\Delta L(T=T_D)}{s} \simeq 7 \times 10^{-3} \, \frac{b_0}{m_N} \,,
\end{align}
for Yukawa couplings $y = \mathcal O(10^{-5})$ and $m_N \simeq 10^5$~GeV, where $s$ is the entropy density of the universe. The numerical coefficient in the right-hand side of \eqref{dltd} depends on the approximation scheme used in the computations. Requiring $\Delta L$ to be the one observed in the Universe (that is of the same order as the observed baryon asymmetry~\cite{Planck}, $\Delta B/s \sim \mathcal O(10^{-10})$ for temperatures $T > 1 ~{\rm GeV}$), we find $b_0 \simeq 10^{-8} \, m_N = \mathcal O({\rm 1~MeV})$.

In \cite{Sarkar3}, the constant $b_0$ analysis has been extended to the case of backgrounds  that depend mildly on the temperature $T$:
\begin{align}\label{bT}
b_0 (T) \sim T^3\,, 
\end{align}
which for the short period of leptogenesis can be assumed approximately constant, so that an analytic treatment is also possible in that case, yielding a lepton asymmetry at decoupling $T_D \simeq m_N$~\cite{Sarkar3}:
\begin{align}\label{dltd2}
\frac{\Delta L (T=T_D)}{s} \simeq  q\, \frac{\mathcal F}{m_N} \,,
\end{align}
where we have set for the $T$-dependent background $b_0(x) \simeq \mathcal F \, x^{-3}$, $x \equiv m_N/T$.
The coefficient $q$ expresses theoretical uncertainties in the approximation methods used, and in particular 
on the expansion point $x_P$ of the appropriate Pad\`e-approximant expansion used in the evaluation~\cite{Sarkar3}. This yields $b_0(T_B)$ in the range: 
\begin{align}\label{b0kev}
b_0(T_D)/{\rm keV} \in [0.36 \, , 0.74]\,,
\end{align}
which produces phenomenologically relevant lepton asymmetries.  The above lepton asymmetries are communicated to the baryon sector via appropriate sphaleron processes in the SM sector, which preserve the difference of Baryon minus Lepton (B-L) number~\cite{Sarkar4}.

Let us now adapt the above mechanism to our stringy RVM model~\cite{bms1,bms2,ms1,ms2}. We have seen that in the presence of gravitational anomaly condensates, induced by condensation of, e.g., primordial GW, there is an induced LV axion background \eqref{axionbackgr}, which survives undiluted until the end of the inflationary period and the onset of radiation. 
In the context of the stringy RVM cosmology, Abelian and non-Abelian gauge fields of the SM sector, and chiral fermionic matter, are all generated at the end of the RVM inflation, as a result of the decay of the metastable RVM vacuum~\cite{Lima1,Lima2}. In the presence of chiral fermions, the r\^ole of the KR field $H_{\mu\nu\rho}$ as torsion implies that its dual KR axion $b(x)$ will couple to the divergence of the axial fermion current $J^{5\mu}$, universally for all fermion species, including RHN if the model contains such fields.
Specifically, the pertinent effective action will read~\cite{bms1}:
\begin{align}\label{sea6}
&S^{\rm eff} =\; \int d^{4}x\sqrt{-g}\Big[ -\dfrac{1}{2\kappa^{2}}\, R + \frac{1}{2}\, \partial_\mu b \, \partial^\mu b -  \sqrt{\frac{2}{3}}\,
\frac{\alpha^\prime}{96\, \kappa} \, \partial_\mu b(x) \, {\mathcal K}^\mu
\Big]  \nonumber \\
&+ S_{\rm Dirac~or~Majorana}^{Free} + \int d^{4}x\sqrt{-g}\, \Big[\Big( {\mathcal F}_\mu + \frac{\alpha^\prime}{2\, \kappa} \, \sqrt{\frac{3}{2}} \, \partial_{\mu}b \Big)\, J^{5\mu}    - \dfrac{3\alpha^{\prime\, 2}}{16 \, \kappa^2}\,J^{5}_{\mu}J^{5\mu} \Big] + \dots,
\end{align}
where $J^{5 \mu}~\equiv~\sum_{i={\rm fermion~species}} \, \overline \psi \gamma^5 \, \gamma^\mu \, \psi $ denotes  the axial fermion current , ${\mathcal F}^d  =   \varepsilon^{abcd} \, e_{b\lambda} \,  \partial_a \, e^\lambda_{\,\,c}$, with $e^\mu_{\,\,c}$ the vielbeins (with Latin indices pertaining to the tangent space of the spacetime manifold at a given point, in a standard notation), 
$S_{\rm Dirac~or~Majorana}^{Free}$ denotes the free-fermion kinetic terms, 
and the $\dots$ in (\ref{sea6}) indicate gauge field kinetic terms, as well as terms of higher order in derivatives. The action \eqref{sea6} is valid for both Dirac or Majorana fermions. The interaction $\partial_\mu b \, J^{5\mu}$ yields, upon partial integration, the aforementioned coupling of the axion $b$ to the divergence of the axial current, and is due to the presence of the torsion on the covariant gravitational derivative for fermions~\cite{torsion,torsion2}. The reader should notice the repulsive four-fermion interactions 
$ - \dfrac{3\alpha^{\prime\, 2}}{16 \, \kappa^2}\,J^{5}_{\mu}J^{5\mu}$ in the action \eqref{sea6}, which is a characteristic feature of the Einstein-Cartan theory of fermionic torsion.

If the axial current is conserved, then the primordial anomaly terms involving the topological current $\mathcal K^\mu$ would remain in the radiation era, and in case there is condensation of GW, one would still obtain a LV background \eqref{axionbackgr} for the KR axion well into the radiation era, which could lead to a constant $b_0$ leptogenesis, according to the mechanism described above ({\it cf.} \eqref{dltd}).

On the other hand, in case there are chiral anomalies, leading to the non-conservation of the axial current, $J_{\quad;\mu}^{5\mu} \ne 0$, there could be a cancellation~\cite{bms1} of the primordial gravitational anomaly $\mathcal K^\mu$-terms in \eqref{sea6} by the gravitational anomalies generated by the chiral fermionic matter, in the sense of:
\begin{align}
   \label{anom2}
   \partial_\mu \Big[\sqrt{-g}\, \Big(  \sqrt{\frac{3}{8}} \kappa\, J^{5\mu}  -  \sqrt{\frac{2}{3}}\,
\frac{\kappa}{96} \, {\mathcal K}^\mu  \Big) \Big] \! & =\!   \sqrt{\frac{3}{8}}\, \frac{\alpha^\prime}{\kappa}\,  \frac{e^2}{8\pi^2}  \, \sqrt{-g}\,  {F}^{\mu\nu}\,  \widetilde{F}_{\mu\nu}  \nonumber \\ &+  \sqrt{\frac{3}{8}} \,\frac{\alpha^\prime}{\kappa}\, \frac{\alpha_s}{8\pi}\, \sqrt{-g} \, G_{\mu\nu}^a \, \widetilde G^{a\mu\nu} \,,
\end{align} 
where $F_{\mu\nu}$ denotes the electromagnetic U(1) Maxwell tensor, and $G_{\mu\nu}^a$, $a=1, \dots 8$ is the gluon tensor associated with the strong interactions with (squared) coupling $\alpha_s = g_s^2/(4\pi)$, and the 
$\widetilde{(\dots)}$ denotes the corresponding duals, as usual.

Upon assuming  \eqref{anom2}, the equation of motion of the KR axion field reads: 
\begin{align}\label{ffgg}
\frac{1}{\sqrt{-g}} \partial_\mu\Big(\sqrt{-g} \, \partial^\mu b\Big) =  - \sqrt{\frac{3}{8}} \, \frac{\alpha^\prime}{\kappa}\,  \frac{e^2}{8\pi^2}\,  {F}^{\mu\nu}\,  \widetilde{F}_{\mu\nu}  -  \sqrt{\frac{3}{8}} \,\frac{\alpha^\prime}{\kappa}\, \frac{\alpha_s}{8\pi}\, G_{\mu\nu}^a \, \widetilde G^{a\mu\nu} \,.
\end{align}
If the terms on the right-hand-side of \eqref{ffgg} are zero at the early radiation era~\cite{ms1}, above the electroweak symmetry breaking, e.g. in the temperature range of the leptogenesis scenarios of \cite{Sarkar3}, $T \sim 10^5~{\rm GeV}$, then one obtains from \eqref{ffgg}, 
\begin{align}\label{bt3}
\dot b \sim T^3\, , 
\end{align}
as a result of the temperature 
dependence of the Universe scale factor $a (T) \sim 1/T$ during the radiation era, of interest to us here. Thus one faces the scenario for leptogenesis of \cite{Sarkar3}, based on axial backgrounds of the form \eqref{bT}, with $b_0 = \dot b$, and the associated lepton asymmetry is given by \eqref{dltd2}, \eqref{b0kev}. 

We remark at this point that for the above scenarios of Leptogenesis to be valid, it is essential that the RHN acquires a non zero mass during the leptogenesis epoch. In our approach we assumed that this is possible, without specifying the detailed mechanism by means of which such a mass is generated. One appealing scenario is the radiative mechanism  for RHN Majorana mass generation of \cite{pilaftsis}, which makes use of the CS gravitational anomaly itself.

Non zero chiral anomalies may appear at much lower temperatures, for instance gluon terms in \eqref{ffgg} may appear due to QCD instanton non-perturbative effects at the QCD epoch, leading to 
the generation of non-perturbative periodic potentials, and thus a mass,  for the KR axion, which in this way may play the r\^ole of a component of DM. For even lower temperatures, one may have the generation of U(1) chiral anomalies, which in turn last until the current era, affecting the temperature evolution of the $b_0(T)= \dot b(T)$ background~\cite{bms1}. Given the connection ({\it cf.} \eqref{dualHb}) of $\dot b$ with the KR totally antisymmetric torsion,  $\dot b \propto \epsilon^{ijk}\, H_{ijk}$, $i,j,k=1,2,3$, one thus obtains a geometric origin of the DM~\cite{MavPhil}. It should also be stressed that the evolution equation of $b_0(T)$ is such that today, $b_0$  is well within the allowed torsion limits~\cite{bms1}, namely, if chiral matter effects are assumed present in the late Universe, then~\cite{bms1}: 
\begin{align}\label{today}
b_0^{\rm today}= \mathcal O(10^{-34})~{\rm eV}\,, 
\end{align}
otherwise it is much more suppressed. The current bounds for $b_0$ are~\cite{smebounds} $b_0 < 10^{-2}$~eV, while the spatial component must satisfy $|b_i| < 10^{-22}$~eV, $i=1,2,3$. The bound \eqref{today} is compatible with $b_i$, even if we take into account the relative velocity of an observer on Earth relative to the cosmic microwave background (CMB) frame~\cite{bms1}.

We conclude this section by remarking that condensates of chiral U(1) anomalies~\cite{bms1} (or even of the combination of U(1) and non-Abelian chiral anomalies, as appears on the right-hand side of \eqref{anom2}) can lead to the appearance of a dark energy component of the Universe at late eras. Such condensates would necessarily be metastable if we want our low-energy model to be embeddable into string theory or perhaps into any other UV complete theory of QG. A  detailed mechanism for the appearance of such a dark energy, which at late eras evolves so slowly with the cosmic time so that it approximates a cosmological constant, in agreement with the plethora of the available data, is still pending within a string theory framework. Nonetheless it seems plausible. At this point it should be remarked that when computing the dark energy of any cosmological model, the back reaction of a quantum field theoretic matter onto the space time may not be ignored, and it is possible that renormalization group effects on the stress-energy tensor of the theory  play a non trivial r\^ole~\cite{Solaqft1,Solaqft2,Solaqft3}. Thus, within our framework of generating dark energy from condensates of gauge fields in the SM sector of our string-inspired RVM cosmology, such effects should be taken properly into account. 

\section{Conclusions and Outlook}\label{sec:concl}
\vspace{0.2cm} 
 
In this talk, we have discussed a cosmological model inspired from string theory, which however was characterised by gravitational anomalies at early epochs. The model is consistent with general covariance, but has interesting and unconventional consequences for the physics of inflation and post inflationary evolution of the Universe. In particular, condensation of primordial GW implies an RVM type inflation, which doies not require external inflaton fields. The inflation arises as a result of the non-linear terms quartic in the Hubble parameter, that characterise the energy density of the model. The anomaly condensates also result in LV and CPTV axion backgrounds, which lead to unconventional Leptogenesis during the radiation era. 

In this framework, the dark sector of the Universe acquires a geometric origin, given the aforementioned association of the dark energy with condensates of anomalies, which could characterise the model beyond the RVM inflationary era, even providing (through condensates of chiral anomalies of gauge type in the modern era) a candidate for a present-era metastable de Sitter space time. Moreover, the association of the gravitational KR axion with the totally antisymmetric torsion,  implies, in case the KR axion develops a non-perturbative mass 
through QCD instanton effects associated with chiral anomalies of the gluon field, a potential geometric origin of dark matter, a component of which could be played by such a massive KR axion.

The modern-era phenomenology of the stringy RVM model is not dissimilar to that of the conventional RVM model. The latter is known to exhibit observable in principle deviations from $\Lambda$CDM~\cite{phenoRVM1,phenoRVM2}, including the potential alleviation of the cosmological data tensions that seem to characterise the modern era of the Universe~\cite{bothRVM}.
Nonetheless, the stringy RVM model includes $H^2\, {\rm ln}H$ corrections in the modern-epoch energy density, as a result of integrating out graviton modes, thus a pure QG effect. Such terms, but not QG in origin, also seem to arise in the renormalisation of quantum fields in the curved spacetime of an expanding Universe as observed recently in \cite{Solaqft1,Solaqft2,Solaqft3}, but there appear to be distinguishing features between those conventional quantum field theory models and the stringy RVM~\cite{gvms}.
Although such non-polynomial-in-$H$ terms do not affect the inflationary scenarios,  nevertheless their detailed phenomenology in the modern era, especially with regards to the data tension alleviation, need yet to be explored. We hope to be able to report soon on this issue.

\section*{Acknowledgments}
\vspace{0.2cm}

I would like to thank  H-T. Elze and the organising committee of the 2022 edition of the DICE series of conferences for the invitation for a plenary talk and for organising such a high level and thought provoking event, which constituted another valuable addition to this excellent conference series.
The work of N.E.M. is supported in part by the UK Science and Technology Facilities research Council (STFC) under the research grant ST/T000759/1. N.E.M.  also acknowledges participation in the COST Association Action CA18108 ``{\it Quantum Gravity Phenomenology in the Multimessenger Approach (QG-MM)}''.

\section*{References}
\vspace{0.1cm}

\bibliography{DICE-bibliography_Mavromatos}

\providecommand{\newblock}{}
\begin{thebibliography}{10}
\expandafter\ifx\csname url\endcsname\relax
  \def\url#1{{\tt #1}}\fi
\expandafter\ifx\csname urlprefix\endcsname\relax\def\urlprefix{URL }\fi
\providecommand{\eprint}[2][]{\url{#2}}

\bibitem{Planck}
Aghanim N {\em et~al.\/} (Planck) 2020 {\em Astron. Astrophys.\/} {\bf 641} A6
  [Erratum: Astron.Astrophys. 652, C4 (2021)] (\textit{Preprint}
  \eprint{1807.06209})

\bibitem{Swamp1}
Ooguri H, Palti E, Shiu G and Vafa C 2019 {\em Phys. Lett. B\/} {\bf 788}
  180--184 (\textit{Preprint} \eprint{1810.05506})

\bibitem{Swamp2}
Obied G, Ooguri H, Spodyneiko L and Vafa C 2018  (\textit{Preprint}
  \eprint{1806.08362})

\bibitem{Swamp3}
Garg S~K and Krishnan C 2019 {\em JHEP\/} {\bf 11} 075 (\textit{Preprint}
  \eprint{1807.05193})

\bibitem{Palti}
Palti E 2019 {\em Fortsch. Phys.\/} {\bf 67} 1900037 (\textit{Preprint}
  \eprint{1903.06239})

\bibitem{tensions}
Verde L, Treu T and Riess A~G 2019 {\em Nature Astron.\/} {\bf 3} 891
  (\textit{Preprint} \eprint{1907.10625})

\bibitem{Freedman}
Freedman W~L 2017 {\em Nature Astron.\/} {\bf 1} 0121, and references therein
  (\textit{Preprint} \eprint{1706.02739})

\bibitem{DiValentino}
Di~Valentino E, Mena O, Pan S, Visinelli L, Yang W, Melchiorri A, Mota D~F,
  Riess A~G and Silk J 2021 {\em Class. Quant. Grav.\/} {\bf 38} 153001, and
  references therein (\textit{Preprint} \eprint{2103.01183})

\bibitem{bothRVM}
Sol\`a~Peracaula J, G\'omez-Valent A, de~Cruz~Perez J and Moreno-Pulido C 2021
  {\em EPL\/} {\bf 134} 19001 (\textit{Preprint} \eprint{2102.12758})

\bibitem{Lima1}
Lima J~A~S, Basilakos S and Sol\`a J 2013 {\em Mon. Not. Roy. Astron. Soc.\/}
  {\bf 431} 923--929 (\textit{Preprint} \eprint{1209.2802})

\bibitem{Lima2}
Perico E~L~D, Lima J~A~S, Basilakos S and Sol\`a J 2013 {\em Phys. Rev. D\/}
  {\bf 88} 063531 (\textit{Preprint} \eprint{1306.0591})

\bibitem{Lima3}
Basilakos S, Lima J~A~S and Sol\`a J 2013 {\em Int. J. Mod. Phys. D\/} {\bf 22}
  1342008 (\textit{Preprint} \eprint{1307.6251})

\bibitem{Limatherm1}
Lima J~A~S, Basilakos S and Sol\`a J 2015 {\em Gen. Rel. Grav.\/} {\bf 47} 40
  (\textit{Preprint} \eprint{1412.5196})

\bibitem{Limatherm2}
Lima J~A~S, Basilakos S and Sol\`a J 2016 {\em Eur. Phys. J. C\/} {\bf 76} 228
  (\textit{Preprint} \eprint{1509.00163})

\bibitem{Solatherm}
Sol\`a~Peracaula J and Yu H 2020 {\em Gen. Rel. Grav.\/} {\bf 52} 17
  (\textit{Preprint} \eprint{1910.01638})

\bibitem{Sola1}
Shapiro I~L and Sol\`a J 2009 {\em Phys. Lett. B\/} {\bf 682} 105--113
  (\textit{Preprint} \eprint{0910.4925})

\bibitem{Sola2}
Shapiro I~L and Sol\`a J 2000 {\em Phys. Lett. B\/} {\bf 475} 236--246
  (\textit{Preprint} \eprint{hep-ph/9910462})

\bibitem{Sola3}
Shapiro I~L and Sol\`a J 2002 {\em JHEP\/} {\bf 02} 006 (\textit{Preprint}
  \eprint{hep-th/0012227})

\bibitem{Sola4}
Sol\`a J 2008 {\em J. Phys. A\/} {\bf 41} 164066 (\textit{Preprint}
  \eprint{0710.4151})

\bibitem{bms1}
Basilakos S, Mavromatos N~E and Sol\`a~Peracaula J 2020 {\em Phys. Rev. D\/}
  {\bf 101} 045001 (\textit{Preprint} \eprint{1907.04890})

\bibitem{bms2}
Basilakos S, Mavromatos N~E and Sol\`a~Peracaula J 2020 {\em Phys. Lett. B\/}
  {\bf 803} 135342 (\textit{Preprint} \eprint{2001.03465})

\bibitem{ms1}
Mavromatos N~E and Sol\`a~Peracaula J 2021 {\em Eur. Phys. J. ST\/} {\bf 230}
  2077--2110 (\textit{Preprint} \eprint{2012.07971})

\bibitem{ms2}
Mavromatos N~E and Sol\`a~Peracaula J 2021 {\em Eur. Phys. J. Plus\/} {\bf 136}
  1152 (\textit{Preprint} \eprint{2105.02659})

\bibitem{scatt1}
Hellerman S, Kaloper N and Susskind L 2001 {\em JHEP\/} {\bf 06} 003
  (\textit{Preprint} \eprint{hep-th/0104180})

\bibitem{scatt2}
Fischler W, Kashani-Poor A, McNees R and Paban S 2001 {\em JHEP\/} {\bf 07} 003
  (\textit{Preprint} \eprint{hep-th/0104181})

\bibitem{Solaqft1}
Moreno-Pulido C and Sol\`a~Peracaula J 2022 {\em Eur. Phys. J. C\/} {\bf 82}
  551 (\textit{Preprint} \eprint{2201.05827})

\bibitem{Solaqft2}
Moreno-Pulido C and Sol\`a J 2020 {\em Eur. Phys. J. C\/} {\bf 80} 692
  (\textit{Preprint} \eprint{2005.03164})

\bibitem{Solaqft3}
Moreno-Pulido C and Sol\`a~Peracaula J 2022 {\em Eur. Phys. J. C\/} {\bf 82}
  1137 (\textit{Preprint} \eprint{2207.07111})

\bibitem{Houston1}
Alexandre J, Houston N and Mavromatos N~E 2013 {\em Phys. Rev. D\/} {\bf 88}
  125017 (\textit{Preprint} \eprint{1310.4122})

\bibitem{Houston2}
Alexandre J, Houston N and Mavromatos N~E 2015 {\em Int. J. Mod. Phys. D\/}
  {\bf 24} 1541004 (\textit{Preprint} \eprint{1409.3183})

\bibitem{bmssugra}
Basilakos S, Mavromatos N~E and Sol\`a J 2016 {\em Universe\/} {\bf 2} 14
  (\textit{Preprint} \eprint{1505.04434})

\bibitem{gvms}
Gomez-V\`alent A, Mavromatos N~E and Sol\`a~Peracaula J  , in progress.

\bibitem{Stamou}
Mavromatos N~E, Spanos V~C and Stamou I~D 2022 {\em Phys. Rev. D\/} {\bf 106}
  063532 (\textit{Preprint} \eprint{2206.07963})

\bibitem{LISA}
Berti E, Cardoso V and Will C~M 2006 {\em Phys. Rev. D\/} {\bf 73} 064030
  (\textit{Preprint} \eprint{gr-qc/0512160})

\bibitem{Fumagalli}
Fumagalli J, Pieroni M, Renaux-Petel S and Witkowski L~T 2022 {\em JCAP\/} {\bf
  07} 020 (\textit{Preprint} \eprint{2112.06903})

\bibitem{string}
Green M~B, Schwarz J~H and Witten E 2012 {\em {Superstring Theory Vol. 2}:
  {25th Anniversary Edition}\/} Cambridge Monographs on Mathematical Physics
  (Cambridge University Press) ISBN 978-1-139-53478-9, 978-1-107-02913-2

\bibitem{torsion}
Hehl F~W, Von Der~Heyde P, Kerlick G~D and Nester J~M 1976 {\em Rev. Mod.
  Phys.\/} {\bf 48} 393--416

\bibitem{gs}
Green M~B and Schwarz J~H 1984 {\em Phys. Lett. B\/} {\bf 149} 117--122

\bibitem{kaloper}
Duncan M~J, Kaloper N and Olive K~A 1992 {\em Nucl. Phys. B\/} {\bf 387}
  215--235

\bibitem{svrcek}
Svrcek P and Witten E 2006 {\em JHEP\/} {\bf 06} 051 (\textit{Preprint}
  \eprint{hep-th/0605206})

\bibitem{jackiw}
Jackiw R and Pi S~Y 2003 {\em Phys. Rev. D\/} {\bf 68} 104012
  (\textit{Preprint} \eprint{gr-qc/0308071})

\bibitem{yunes}
Alexander S and Yunes N 2009 {\em Phys. Rept.\/} {\bf 480} 1--55
  (\textit{Preprint} \eprint{0907.2562})

\bibitem{alexander}
Alexander S~H~S, Peskin M~E and Sheikh-Jabbari M~M 2006 {\em Phys. Rev.
  Lett.\/} {\bf 96} 081301 (\textit{Preprint} \eprint{hep-th/0403069})

\bibitem{lyth}
lyth D~H, Quimbay C and Rodriguez Y 2005 {\em JHEP\/} {\bf 03} 016
  (\textit{Preprint} \eprint{hep-th/0501153})

\bibitem{ellisinfl}
Ellis J and Mavromatos N~E 2013 {\em Phys. Rev. D\/} {\bf 88} 085029
  (\textit{Preprint} \eprint{1308.1906})

\bibitem{MavLV}
Mavromatos N~E 2022 {\em {740. WE-Heraeus-Seminar}: {Experimental Tests and
  Signatures of Modified and Quantum Gravity Workshop}\/} (Chapter in the
  Springer Book Modified and Quantum Gravity - From theory to experimental
  searches on all scales - WEH 740, (C. Laemmerzahl and C. Pfeifer eds)
  (\textit{Preprint} \eprint{2205.07044})

\bibitem{kanti}
Kanti P, Mavromatos N~E, Rizos J, Tamvakis K and Winstanley E 1996 {\em Phys.
  Rev. D\/} {\bf 54} 5049--5058 (\textit{Preprint} \eprint{hep-th/9511071})

\bibitem{Sarkar1}
de~Cesare M, Mavromatos N~E and Sarkar S 2015 {\em Eur. Phys. J. C\/} {\bf 75}
  514 (\textit{Preprint} \eprint{1412.7077})

\bibitem{Sarkar2}
Bossingham T, Mavromatos N~E and Sarkar S 2018 {\em Eur. Phys. J. C\/} {\bf 78}
  113 (\textit{Preprint} \eprint{1712.03312})

\bibitem{Sarkar3}
Bossingham T, Mavromatos N~E and Sarkar S 2019 {\em Eur. Phys. J. C\/} {\bf 79}
  50 (\textit{Preprint} \eprint{1810.13384})

\bibitem{Sarkar4}
Mavromatos N~E and Sarkar S 2020 {\em Eur. Phys. J. C\/} {\bf 80} 558
  (\textit{Preprint} \eprint{2004.10628})

\bibitem{axiverse}
Arvanitaki A, Dimopoulos S, Dubovsky S, Kaloper N and March-Russell J 2010 {\em
  Phys. Rev. D\/} {\bf 81} 123530 (\textit{Preprint} \eprint{0905.4720})

\bibitem{silver}
McAllister L, Silverstein E and Westphal A 2010 {\em Phys. Rev. D\/} {\bf 82}
  046003 (\textit{Preprint} \eprint{0808.0706})

\bibitem{tetradis1}
Kefala K, Kodaxis G~P, Stamou I~D and Tetradis N 2021 {\em Phys. Rev. D\/} {\bf
  104} 023506 (\textit{Preprint} \eprint{2010.12483})

\bibitem{tetradis2}
Dalianis I, Kodaxis G~P, Stamou I~D, Tetradis N and Tsigkas-Kouvelis A 2021
  {\em Phys. Rev. D\/} {\bf 104} 103510 (\textit{Preprint} \eprint{2106.02467})

\bibitem{Fumagalli2}
Fumagalli J, Renaux-Petel S and Witkowski L~T 2021 {\em JCAP\/} {\bf 08} 030
  (\textit{Preprint} \eprint{2012.02761})

\bibitem{Fumagalli3}
Fumagalli J, Renaux-Petel S~e and Witkowski L~T 2021 {\em JCAP\/} {\bf 08} 059
  (\textit{Preprint} \eprint{2105.06481})

\bibitem{Carr1}
Carr B, Kuhnel F and Sandstad M 2016 {\em Phys. Rev. D\/} {\bf 94} 083504
  (\textit{Preprint} \eprint{1607.06077})

\bibitem{Carr2}
Carr B and Kuhnel F 2020 {\em Ann. Rev. Nucl. Part. Sci.\/} {\bf 70} 355--394
  (\textit{Preprint} \eprint{2006.02838})

\bibitem{pBHDM1}
Clesse S and Garc\'\i{}a-Bellido J 2017 {\em Phys. Dark Univ.\/} {\bf 15}
  142--147 (\textit{Preprint} \eprint{1603.05234})

\bibitem{pBHDM2}
Garc\'\i{}a-Bellido J, Carr B and Clesse S 2021 {\em Universe\/} {\bf 8} 12
  (\textit{Preprint} \eprint{1904.11482})

\bibitem{sasaki}
Zhou Z, Jiang J, Cai Y~F, Sasaki M and Pi S 2020 {\em Phys. Rev. D\/} {\bf 102}
  103527 (\textit{Preprint} \eprint{2010.03537})

\bibitem{seesaw1}
Minkowski P 1977 {\em Phys. Lett. B\/} {\bf 67} 421--428

\bibitem{Yanagida}
Yanagida T 1979 {\em Conf. Proc. C\/} {\bf 7902131} 95--99

\bibitem{GellMann}
Gell-Mann M, Ramond P and Slansky R 1979 {\em Conf. Proc. C\/} {\bf 790927}
  315--321 (\textit{Preprint} \eprint{1306.4669})

\bibitem{seesaw2}
Schechter J and Valle J~W~F 1980 {\em Phys. Rev. D\/} {\bf 22} 2227

\bibitem{seesaw3}
Mohapatra R~N and Senjanovic G 1981 {\em Phys. Rev. D\/} {\bf 23} 165

\bibitem{seesaw4}
Lazarides G, Shafi Q and Wetterich C 1981 {\em Nucl. Phys. B\/} {\bf 181}
  287--300

\bibitem{sme1}
Colladay D and Kostelecky V~A 1998 {\em Phys. Rev. D\/} {\bf 58} 116002
  (\textit{Preprint} \eprint{hep-ph/9809521})

\bibitem{smebounds}
Kostelecky V~A and Russell N 2011 {\em Rev. Mod. Phys.\/} {\bf 83} 11--31
  (\textit{Preprint} \eprint{0801.0287})

\bibitem{torsion2}
Shapiro I~L 2002 {\em Phys. Rept.\/} {\bf 357} 113 (\textit{Preprint}
  \eprint{hep-th/0103093})

\bibitem{pilaftsis}
Mavromatos N~E and Pilaftsis A 2012 {\em Phys. Rev. D\/} {\bf 86} 124038
  (\textit{Preprint} \eprint{1209.6387})

\bibitem{MavPhil}
Mavromatos N~E 2022 {\em Phil. Trans. A. Math. Phys. Eng. Sci.\/} {\bf 380}
  20210188 (\textit{Preprint} \eprint{2108.02152})

\bibitem{phenoRVM1}
Sol\`a~Peracaula J, de~Cruz~P\'erez J and Gomez-Valent A 2018 {\em Mon. Not.
  Roy. Astron. Soc.\/} {\bf 478} 4357--4373 (\textit{Preprint}
  \eprint{1703.08218})

\bibitem{phenoRVM2}
Sol\`a~Peracaula J, Gomez-Valent A and de~Cruz~P\'erez J 2019 {\em Phys. Dark
  Univ.\/} {\bf 25} 100311 (\textit{Preprint} \eprint{1811.03505})

\end{thebibliography}

\end{document}